\documentclass[aps,prd,showpacs,preprintnumbers,nofootinbib,twocolumn]{revtex4}
\usepackage{bm}
\usepackage{latexsym}
\usepackage{dcolumn}
\usepackage{amsmath,amsfonts,amssymb}
\usepackage{graphicx,epsfig}
\usepackage{psfrag}
\usepackage{amsthm}

\def\be {\begin{equation}}
\def\ee {\end{equation}}
\def\bea {\begin{eqnarray}}
\def\eea {\end{eqnarray}}
\def\bc {\begin{center}}
\def\ec {\end{center}}
\def\bfg {\begin{figure}}
\def\efg {\end{figure}}
\def\bi {\begin{itemize}}
\def\ei {\end{itemize}}
\def\nn {\nonumber}

\def\la {\label}
\def\le {\left}
\def\ri {\right}
\def\pa {\partial}
\def\fr {\frac}
\def\sq {\sqrt}

\def\a  {\alpha}
\def\b  {\beta}
\def\c  {\gamma}

\def\d  {\delta}

\def\e  {\epsilon}
\def\f {\phi}

\def\k  {\kappa}
\def\l  {\lambda}
\def\L  {\Lambda}
\def\m  {\mu}
\def\n  {\nu}

\def\O  {\Omega}
\def\p  {\pi}
\def\r  {\rho}
\def\th {\theta}
\def\s {\sigma}
\def\t  {\tau}
\def\vph {\varphi}

\def\cA {\mathcal A}
\def\cN {\mathcal N}
\def\cH {\mathcal H}
\def\cE {\mathcal E}
\def\xh {\hat{x}}

\def\kt {\tilde{\kappa}}
\def\et {\tilde{\epsilon}}
\def\beq{\begin{equation}}
\def\eeq{\end{equation}}
\def\br{\begin{eqnarray}}
\def\er{\end{eqnarray}}
\newcommand{\eel}[1] {\label{#1}\end{equation}}

\newcommand{\bdm}{\begin{displaymath}}
\newcommand{\edm}{\end{displaymath}}

\begin{document}

\title{Power-law corrections to entanglement entropy of horizons}

\author{Saurya Das$^1$}\email[email: ]{saurya.das@uleth.ca}
\author{S. Shankaranarayanan$^2$}\email[email: ]{shanki@aei.mpg.de}
\author{Sourav Sur$^{1}$}\email[email: ]{sourav.sur@uleth.ca}

\affiliation{$^1$~Dept. of Physics,
University of Lethbridge, 4401 University Drive, 
Lethbridge, Alberta, Canada T1K 3M4 \\ 
$^2$~Max-Planck-Institut f\"ur Gravitationphysik, 
Am M\"uhlenberg 1, D-14476 Potsdam, Germany}

\begin{abstract}
We re-examine the idea that the origin of black-hole entropy 
may lie in the entanglement of quantum fields between inside and
outside of the horizon. Motivated by the observation that certain
modes of gravitational fluctuations in a black-hole background behave
as scalar fields, we compute the entanglement entropy of such a field,
by tracing over its degrees of freedom inside a sphere. We show that 
while this entropy is proportional to the area of the sphere when the 
field is in its ground state, a correction term proportional to a 
fractional power of area results when the field is in a superposition 
of ground and excited states. The area law is thus recovered for large 
areas. Further, we identify location of the degrees of freedom that give 
rise to the above entropy.
\end{abstract}

\maketitle


\section{Introduction \la{intro}}

Almost four decades ago, Bekenstein realized that if the second law of
thermodynamics is not to be violated in the presence of a black hole,
the black-hole must possess an entropy proportional to its horizon area 
\cite{bek}. The significance of this result became clear with Hawking's 
demonstration of black-hole thermal radiation \cite{haw}. Hawking showed 
that quantum effects in the background of a body collapsing to a 
Schwarzschild black-hole of mass $M$, will lead, at late times, to a 
radiation of particles in all modes of the quantum field, with a 
characteristic thermal spectrum at a temperature equal to
\beq
T_{\rm H} = \le(\frac{\hbar c}{k_{_{B}}} \ri) \frac{\kappa}{2 \pi} = 
\le(\frac{\hbar c^3}{G k_{_{B}}}\ri) \frac{1}{8 \pi M}\, ,
\eeq
where $\kappa$ is the surface gravity of the black-hole, $G$ is the
four-dimensional Newton's constant and $k_{_{B}}$ is the Boltzmann
constant.  Since the Hawking temperature fixes the factor of
proportionality between temperature and surface gravity, one finds the
Bekenstein-Hawking area law
\beq
\la{temp1}
S_{_{\rm BH}} = \le(\frac{k_{_{B}}}{4}\ri) 
\frac{\cA_{\rm H}}{\ell_{_{\rm Pl}}^2} \quad
\mbox{where} \quad {\ell_{_{\rm Pl}}}=\sqrt{\frac{G\hbar}{c^3}} \, ,
\eeq
$\cA_{\rm H}$ is the horizon area and ${\ell_{_{\rm Pl}}}$ is
the four-dimensional Planck length.  All known black-holes in $n (>
2)$ space-time dimensions satisfy the area law (AL).

The universality of the AL has raised some important questions that
remain unanswered: what is the dynamical mechanism that makes
$S_{_{\rm BH}}$ a universal function, independent of the black-hole's
past history and detailed internal condition? Why is $S_{_{\rm BH}}$
proportional to $\cA_{\rm H}$?  What is the microscopic origin of BH
entropy? Are there corrections to the entropy and if so, how generic
are these corrections? Where are the degrees of freedom, responsible
for the entropy, located? These questions often seem related, which a
correct theory of quantum gravity is expected to address.

Naturally, there has been considerable work attempting to address some
of the above questions (for recent reviews, see Refs.
\cite{wald,entropyrev}). Broadly, there have been two approaches: (i)
associating $S_{_{\rm BH}}$ with fundamental degrees of freedom of a
microscopic theory of quantum gravity \cite{stringsetc} and (ii)
associating $S_{_{\rm BH}}$ with quantum matter fields propagating in
a fixed BH background \cite{bkls,sred,thooft}.

In this work, we focus on the second approach and, in particular, we
attribute $S_{_{\rm BH}}$ to the entanglement of the quantum fields
inside and outside the horizon. We show that it is possible to: (a)
obtain generic power-law corrections to Eq.~(\ref{temp1}) which are
small for large horizon radii, but lead to fractional powers of area
for small horizons, (b) identify the degrees of freedom that give rise
to entanglement entropy and (c) test the robustness of
Eq.~(\ref{temp1}) and its corrections for massive quantum fields.

Consider a quantum scalar field (in a pure state) propagating in the
BH space-time.  For an outside observer, the BH horizon provides a
boundary and (s)he can only have information of the state restricted
outside the horizon. In other words, while the full state of the field
is pure, the state restricted outside the horizon is mixed, which leads
to a non-zero entropy. This entropy, {\it aka} Von Neumann entropy,
can formally be written as:
\be 
\la{entangle}
S ~=~ - k_{_B} \mbox{Tr} \le(\r \ln \r\ri)
\ee
where $\r$ is the mixed (or {\it reduced}) density matrix. The above
definition is the microcanonical definition of entropy will be used
here. Although it is also possible to compute the entanglement entropy 
in the canonical picture \cite{thooft}, its usage is restrictive due 
to the fact that it implicitly assumes positive specific heat. On the 
contrary, however, Schwarzschild BH has a negative specific heat.

About two decades ago, Bombelli {\it et al} \cite{bkls} showed that the 
entanglement entropy of scalar fields is proportional to $\cA_{\rm H}/a^2$, 
where $\cA_{\rm H}$ is the area of the boundary of the region being traced
over (the ``horizon"), and $a$ is the ultraviolet cut-off at the horizon
(equivalently, the lattice spacing, when space is discretized). Identifying 
$a$ with the Planck length ${\ell_{Pl}}$, one recovers the Bekenstein-Hawking 
AL (\ref{temp1}). These results were later reproduced by Srednicki \cite{sred},
where by tracing over the degrees of freedom inside a spherical surface of
radius $R$, he showed that the entanglement entropy
\be
S ~\sim~ \frac{\cA}{a^2} \, ,
\ee
where $\cA = 4 \pi R^2$. Thus, the area law can be considered as a consequence 
of the entanglement of the quantum fields across the horizon\footnote{Although, 
one recovers the area law, the divergence of the entanglement entropy has been 
a puzzle. The reason for the divergence is due to the fact that the boundary 
delineating the region being traced over is sharp \cite{Bek:1994bc}.}. (An 
analytical proof of the area proportionality has recently been given in
Ref. \cite{eisert}.  For an application of entanglement in stringy
black-holes, see Refs. \cite{arom,ryu}.)

Now, why is such a computation involving scalar fields in flat space-time
relevant for black-holes at all? Here, we try to provide at least a plausible
answer to this question: in Appendix \ref{app:sf-mot}, we consider gravitational 
perturbations in a black-hole background, and show that for certain modes of 
these perturbations, the effective action reduces to that of a scalar field. 
Further, in Appendix \ref{app:BH-Ham}, we write the corresponding scalar field 
Hamiltonian in the black-hole background in Lema\^itre coordinates, and show that 
for a {\it fixed} Lema\^itre time, it reduces to that in a flat space-time. Thus 
for time-independent quantities such as entropy, scalar fields of the type 
considered here appear to play an important role.

The computation and the area proportionality of entanglement entropy
by Bombelli {\it et al} \cite{bkls} and Srednicki \cite{sred} is based
on the {\it simplifying assumption} that the quantum field is in the
{\it vacuum (i.e., ground) state} (GS). Recently, two of the authors
(S.D. and S.Sh.) investigated the robustness of the entanglement entropy:
area law relation by considering non-vacuum states \cite{sdshanki,sdshankiES}. 
It was shown that while replacing the vacuum state by {\it generic coherent 
state} (GCS) or a class of {\it squeezed states} (SS) does not affect the AL, 
doing so with a class of {\it n-particle or excited states} (ES) results in 
a significant deviation from the AL. More specifically, if the scalar field 
is in a class of $1$-particle ES, it was shown that the entropy scales as
\be \la{es-al}
S ~\sim~ \le(\frac{\cA}{a^2}\ri)^\m \, ,
\ee
where the power $\m$ is always less than unity, and decreases with the
increase in the number of excitations \cite{sdshankiES} (see the
Appendix (\ref{app:GSES-ent}) for details).  Thus, it was shown that the
entanglement entropy does not always lead to AL and its form crucially 
depends on the choice of the quantum state.

Given the above results, one may draw two distinct conclusions: first
--- that entanglement entropy is not robust --- and {\it reject} it as
a possible source of BH entropy.  Second --- since entanglement
entropy for ES scales as a lower power of area --- it is plausible
that when a generic state (consisting of a superposition of GS and ES)
is considered, corrections to the Bekenstein-Hawking entropy will
emerge.  In order to determine which one is correct, it is imperative
to investigate various generalizations of the scenarios considered in
Refs. \cite{bkls,sred,sdshankiES}. To this end, in this work we calculate 
the entanglement entropy of the {\it mixed --- superposition of vacuum 
and 1-particle -- state} (MS).  We show explicitly that the MS entanglement 
entropy is given by
\be
S = c_0 \le(\frac{\cA}{a^2}\ri) \le[1 + c_1 f(\cA)\ri] ~~
\mbox{where} ~~ f(\cA) = \le(\frac{\cA}{a^2}\ri)^{- \nu}\!\!\!\! ,
\ee
$c_0, c_1$ are constants of order unity, and $\nu$ is a fractional
power which depends on the amount of mixing.  Thus, we show that, for
large horizon area ($\cA \gg a^2$), the contribution of $f(\cA)$ is
negligible and the MS entanglement entropy asymptotically approaches
the GS entropy.  This is significantly different from the $1$-particle
state considered in Ref. \cite{sdshankiES} for which the entropy
always scales as power of area, cf. Eq.(\ref{es-al}), the power being
less than unity. Thus, if black-hole entropy is a consequence of
quantum entanglement, the AL is valid for large horizons, as long as
the quantum field is in a superposition of vacuum and 1-particle
states.

From a physics point of view, we expect power-law corrections to
Bekenstein-Hawking entropy for the following two reasons: (a)
$S_{_{\rm BH}}$ is a semi-classical result and is valid for large
black-holes, i.e., when $r_h \gg \ell_{_{\rm Pl}}$ ($r_h$ is the radius
of the horizon).  It is not clear, whether the AL will be valid for
the small BHs ($r_h \sim \ell_{_{\rm Pl}}$).  (b) There is no reason
to expect that the Bekenstein-Hawking entropy to be the whole answer
for a correct theory of quantum gravity.  For instance, it was shown
by Wald \cite{Wald:1993a} that if one takes into account higher
curvature corrections to the Einstein-Hilbert action, the
Bekenstein-Hawking entropy is the leading term in a series expansion.

As mentioned earlier, in this paper, we also identify the precise
location of the microscopic degrees of freedom (DoF) for the
entanglement entropy of the superposition of vacuum and 1-particle
states \cite{sdshankiDoF}.  We find that the DoF close to the horizon
contribute most to the total entropy.  However, there are small
contributions from the DoF far away from the horizon as well.  These
far-away DoF contributions are least in the case of vacuum state and
increase as the number of excitations and/or the mixing weight of
$1$-particle state with vacuum state increases.  Correspondingly,
deviation from the AL increases as well.  Thus, the corrections to
the AL may, in a way, be attributed to the far-away DoF.

Finally, we investigate the effect of massive fields on entanglement
entropy.  We show that in all cases (vacuum, $1$-particle and
superposed states), the massive scalar field entanglement entropy
($S_m$) contains an exponential damping factor in comparison with the
massless scalar field entanglement entropy ($S_0$).  The Gaussian fits
of the ratio $S_m/S_0$ with the mass (in appropriate units) show that
the exponential factor depends explicitly on the mass squared and
hence falls off rapidly as the mass is increased. However, there is
not much variation of the fitting parameters for the different ---
vacuum, 1-particle and mixed --- states, even when a fairly high
amount of excitation is taken into account. This shows that the mixing
proportions in the GS and ES superposition have little influence on
the ratio $S_m/S_0$. We thus show that the mass overall reduces the
entropy exponentially.

The organization of this paper is as follows: In the next section,
we briefly review the procedure of obtaining the entanglement entropy 
of massless scalar fields in flat space-time.  In Sec. (\ref{ent-gses}), 
we obtain the (reduced) density matrix for the scalar field which is in 
a superposition of GS and $1$-particle ES.  We compute the entanglement 
entropy numerically for such superposition and estimate the corrections 
to the BH area law. In Sec. (\ref{dof}), we locate the scalar field 
degrees of freedom that are responsible for the entanglement entropy 
for the superposition of GS and ES. In Sec. (\ref{ent-massive}), we 
obtain the entanglement entropy for a massive scalar field.  We conclude
with a summary and open questions in Sec. (\ref{conclu}). In Appendix
\ref{app:sf-mot}, we discuss the motivation for considering massless 
or massive scalar field for computing the entanglement entropy, and as
mentioned before, we show that for certain modes of gravitational 
perturbations, the relevant action reduces to that of a scalar field. In 
Appendix \ref{app:BH-Ham}, we obtain the Hamiltonian of a scalar field in 
a general BH space-time. We show that this Hamiltonian in Lema\^itre 
coordinates, and at a fixed Lema\^itre time, reduces to the scalar field 
Hamiltonian in flat space-time. Thus this Hamiltonian is relevant for the 
computation of time-independent quantities such as entropy. In Appendix
\ref{app:GSES-ent} we briefly review the results obtained in the earlier
works \cite{bkls,sred,sdshankiES} for the ground state and 1-particle state.

Before we proceed, a few comments on the notation we use are in order:
The metric is four-dimensional with the signature $(-,+,+,+)$. We use
units with $k_{_{B}} = c = \hbar = 1$ and set $M_{_{\rm Pl}}^2 = 1/(16
\pi G)$.  The quantum field $\varphi$ is a minimally coupled scalar
field.

\section{Entanglement entropy of scalar fields}        
\la{ent-scalar}

In this section, we briefly review the procedure of obtaining entanglement 
entropy for scalar fields propagating in flat space-time. The motivation 
for considering scalar fields for the entanglement entropy computations is 
discussed in Appendix \ref{app:sf-mot}. The relevance of the scalar field 
Hamiltonian in flat space-time for computing entropy in a black-hole
space-time is discussed in Appendix \ref{app:BH-Ham}.

The Hamiltonian of massless scalar field propagating in flat space-time is 
given by Eq. (\ref{ham2}). In order to
obtain 
the entropy, we need to discretize this Hamiltonian on the radial lattice
with lattice spacing $a$. Discretizing the Hamiltonian such that $r
\rightarrow r_i;~ r_{i+1} - r_i = a$ and $L = (N+1) a$ is the infrared
cut-off\footnote{In discretizing the terms containing the
derivatives, one usually adopts the middle-point prescription, i.e.,
the derivative of the form $f(x) d_x[g(x)]$ is replaced by $f_{j +
1/2} [g_{j + 1} - g_j]/a$.}, we get
\bea \label{disc1}
H_{lm} &=& \fr 1 {2a} \sum_{j=1}^N \le[ \p_{lm,j}^2 + 
\le(j + \fr 1 2\ri)^2 \le(\fr{\varphi_{lm,j}}{j} - 
\fr{\varphi_{lm,j+1}}{j+1}\ri)^2 \ri. \nn \\
&+& \le. \fr{l(l+1)}{j^2}~\varphi_{lm,j}^2 
\ri]  \quad , \quad H = \sum_{lm} H_{lm}  ,
\eea
where $\varphi_{lm,j} \equiv \varphi_{lm}(r_j), ~ \p_{lm} 
\equiv \p_{lm,j}(r_j)$ and $[\varphi_{lm,j}, \p_{l'm',j'}] = 
i \d_{l l'}\d_{m m'}\d_{j j'}$. 
Up to the overall factor of $a^{-1}$, Eq.~(\ref{disc1}) is identical to the
Hamiltonian of $N-$coupled harmonic oscillators (HOs):
\be \la{coupledham1}
H ~=~ \fr 1 2 \sum_{i=1}^N p_i^2 
~+~ \fr 1 2 \sum_{i,j=1}^N x_i K_{ij} x_j \, ,
\ee
where the matrix $K_{ij}$ represents the potential energy and the
interaction between the oscillators ($i,j = 1,\dots,N$, the
coordinates $x_i$ replace the field variables $\varphi_{lm}$).
For the Hamiltonian (\ref{disc1}), it is given by:
\bea \la{kij}
K_{ij} &=&  \fr 1 {i^2} \le[l(l+1)~\delta_{ij} + \fr 9 4 ~\d_{i1} 
\d_{j1} + \le( N - \fr 1 2\ri)^2 \d_{iN} \d_{jN} \ri. \nn\\
&&+ \le. \le\{ \le(i + \fr 1 2\ri)^2 + \le(i - \fr 1 2\ri)^2 \ri\} 
\d_{i,j(i\neq 1,N)}\ri] \nn\\ 
&&- \le[\fr{(j + \fr 1 2)^2}{j(j+1)}\ri] \delta_{i,j+1} - 
\le[\fr{(i + \fr 1 2)^2}{i(i+1)}\ri] \delta_{i,j-1} .
\eea
The last two terms denote nearest-neighbour interactions and originate 
from the derivative term in Eq. (\ref{ham2}). The most general eigen-state 
of the Hamiltonian (\ref{coupledham1}) is a product of $N-$HO wave functions:
\be \la{excwavefn1}
\psi (x_1,\dots,x_N) = \prod_{i=1}^N \cN_i ~\cH_{\n_i} 
\le(k_{Di}^{1/4}~{\underbar x}_i \ri)
\exp\le( -\fr 1 2 k_{Di}^{1/2}~{\underbar x}_i^2 \ri), 
\ee
where $\cN_i$ s are the normalization constants given by
\be \la{normconst}
\cN_i = \fr{k_{Di}^{1/4}}{\p^{1/4}~\sqrt{2^{\n_i} \n_i!}} \, ,
\ee
${\underbar x} = Ux$, ($U^TU=I_N$), $x^T = (x_1,\dots,x_N)$, 
${\underbar x}^T = ({\underbar x}_1,\dots,{\underbar x}_N)$,
$K_D \equiv U K U^T$ is a diagonal matrix with elements $k_{Di}$, 
and $\nu_i \, (i=1 \dots N)$ are the indices of the Hermite 
polynomials ($\cH_{\nu}$). The frequencies are ordered such that 
$k_{Di} > k_{Dj}$ for $i > j$. 

Defining the $N \times N$ matrix $\O = U^T K_D^{1/2} U$, such that
$|\O| = |K_D|^{1/2}$, and tracing over first $n$ of the $N$ 
oscillators, one obtains the reduced density matrix:
\bea \la{denmatgen1}
\r \le(t; t'\ri) &=& \int \prod_{i=1}^n dx_i ~ 
\psi(x_1,\dots,x_n; t) ~\psi^\star(x_1,\dots,x_n; t') \nn \\
&=& \int \prod_{i=1}^n dx_i \exp \le[-\fr{x^T \O x} 2\ri] 
\prod_{i=1}^{N} \cN_i \cH_{\n_i} \le(k_{Di}^{1/4} 
{\underbar x}_i\ri) \nn\\
&& \times~ \exp \le[-\fr{x'^T \O x'} 2\ri] \prod_{j=1}^{N} 
\cN_j \cH_{\n_j} \le(k_{Di}^{1/4} {\underbar x}'_i\ri)
\eea
where we now denote: $x^T = (x_1, \dots ,x_n ; t_1, \dots, t_{N-1}) 
= (x_1, \dots, x_n ; t)$, with $t \equiv t_1, \dots, t_{N-n} ~; t_j 
\equiv x_{n+j}, ~j =1, \dots, (N-n)$. It is easy to check that 
$\r^2 \neq \r$, implying that $\r$ is mixed, i.e, although the full 
state is pure, the state obtained by integrating over $n$-HO is mixed. 
Substituting the reduced density matrix (\ref{denmatgen1}) into the 
formal expression (\ref{entangle}) will yield a non-zero (positive) 
entanglement entropy.

It is not possible to obtain a closed form expression for the density
matrix for an arbitrary state (\ref{excwavefn1}). However, in the cases 
where all the HOs are in their GS \cite{sred}, or in the GCS or in a 
class of SS \cite{sdshankiES}, all of which are minimum uncertainty 
states, closed form analytic expressions of $\r(t;t')$, and hence of 
the entropy, can be evaluated exactly and shown to follow the BH AL. 
For the first ES, not a minimum uncertainty state, the entropy computed 
numerically \cite{sdshankiES} is found not to obey the AL.

In the following section, we obtain the entanglement entropy for the
superposition of GS and ES. [For the sake of completeness, we have
briefly discussed the entanglement entropy for the ground and
first-excited states in the Appendix \ref{app:GSES-ent}.]

\section{Entanglement entropy for a Superposition of GS and ES} 
\la{ent-gses}

In this section, we obtain the entanglement entropy for the superposition 
of ground and excited states. (In the following, we denote all relevant 
quantities such as the wave function, density matrix etc by the symbol/suffix 
$0$ for GS and by $1$ for the first ES.)

The discretized scalar field wave function $\psi$ in a MS is a linear 
superposition of the N-HO GS wave function $\psi_0$, Eq. (\ref{gs-wavefn})
[Appendix \ref{app:GSES-ent}], and N-HO (1-particle) ES wave function 
$\psi_1$ [corresponding to one HO in the ES, while the rest $N-1$ in their 
GS, Eq. (\ref{es-wavefn})], i. e.,
\be \la{gses-wavefn}
\psi (\xh; t) ~=~ \le[c_0 ~\psi_0 (\xh; t) ~+~ c_1 ~\psi_1 (\xh; t)\ri] 
\ee
where $\xh \equiv \{x_1, \cdots, x_n\}~$; and as before $t_j \equiv x_{n+j} 
~ (j = 1, \cdots, N-n) ~;~ t \equiv \{t_1, \cdots, t_{N-n}\} = \{x_{n+1}, 
\cdots, x_N\}$. Normalization of $\psi$ requires $c_0^2 + c_1^2 = 1$. Here 
we assume that $c_0$ and $c_1$ are real constants.

Referring to the Appendix \ref{app:GSES-ent} and using Eq. (\ref{es-wavefn}), 
we can write,
\be \la{gses-wavefn1}
\psi (\xh; t) ~=~ \le[c_0 ~+~ c_1 ~ f (\xh; t)\ri] \psi_0 (\xh; t) \, ,
\ee
where,
\be \la{f}
f (\xh; t) ~=~ \sq{2} \a^T K_D^{1/4} U x ~=~ y^T x \, ,
\ee
$\a$ being the expansion coefficient defined in Eq. (\ref{expcoeff}). The $N$ 
dimensional column vector $y$ is given by
\bea \la{y}
y ~=~ \sq{2} U^T K_D^{1/4} \a ~=~ \le( \begin{array}{l} 
{y_A} \\ 
{y_B} 
\end{array} \ri) 
\eea
$y_A$ and $y_B$ are $n$- and $(N-n)$-dimensional column vectors, respectively.

The density matrix is a sum of three terms:
\bea \la{gses-den}
\r (t; t') &=& \int \prod_{i=1}^n dx_i~ \psi (\xh; t) 
\psi^\star (\xh; t') \\
&=& c_0^2 ~\r_0 (t; t') ~+~ c_1^2 ~\r_1 (t; t') ~+~ c_0 c_1 
~\r_2 (t; t') \nn
\eea
where $\r_0 (t; t')$ is the GS density matrix (\ref{gs-den}), $\r_1 (t; t')$ 
is the ES density matrix (\ref{es-den}). It is easy to see that, one can make 
the following identifications of the matrix $\Lambda$ and it's components; and 
the constant $\k$ (see Appendix \ref{app:GSES-ent}), with the column vector $y$ 
and it's components as
\bea \la{identify}
&& \L ~=~ \fr 1 2 y y^T ~=~ \le( \begin{array}{ll} 
{\L_A} & {\L_B} \\
{\L_B^T} & {\L_C} 
\end{array} \ri), \nn\\
&& \L_A = \fr 1 2 y_A y_A^T ~;~ \L_B = \fr 1 2 y_A y_B^T ~;~ 
\L_C = \fr 1 2 y_B y_B^T ~, \nn\\
&& \k ~=~ \mbox{Tr} (\L_A A^{-1}) ~=~ \fr 1 2 y_A^T A^{-1} y_A ~.
\eea
$\r_2$ is the cross-term in the total density matrix $\r$, Eq. (\ref{gses-den}), 
due to the mixing of GS and ES and can be evaluated as follows:
\bea \la{ms-den}
\r_2 (t; t') &=& \int \prod_{i=1}^n dx_i \le[f (\xh; t) + 
f (\xh; t')\ri] \psi_0 (\xh; t) ~ \psi_0^\star (\xh; t') \nn\\
&=& \le(y_B - p\ri)^T \le(t + t'\ri) \r_0 (t; t') \, ,
\eea
where
\be \la{p}
p ~=~ B^T A^{-1} y_A ~,
\ee
is an $(N-n)$-dimensional column vector\footnote{For definitions of matrices
$A, B$ etc. see the Appendix \ref{app:GSES-ent} [Eq. (\ref{Omega})].}.

Using Eqs. (\ref{gs-den}), (\ref{es-den}) and (\ref{ms-den}), the complete MS
density matrix can be written as
\be \la{gses-den1}
\r (t; t') = \le[c_0^2 + c_1^2 \k \le\{1 + u (t; t')\ri\} + c_0 c_1  v (t; t')
\ri] \r_0 (t; t')
\ee
where the functions $u$ and $v$ are defined by
\bea \la{uv}
u (t; t') &=& - ~ \fr{t^T \L_\c t + t'^T \L_\c t'} 2 ~+~ t^T
\L_\b t' \nn\\
v (t; t') &=& \le(y_B - p\ri)^T \le(t + t'\ri). 
\eea
Let us define
\be \la{F}
F (t; t') ~=~ 1 ~+~ \k_1 w (t; t') ~+~ \k_2 v (t; t') ~+~ 
\fr {\k_2^2} 2 v^2 (t; t')
\ee
where
\br 
\la{w}
w (t; t')&=& - ~ \fr{t^T \L_{\c'} t + t'^T \L_{\c'} t'} 2 ~+~ 
t^T \L_{\b'} t'  \nn \\ 
\L_{\b'} &=& \L_\b ~-~ 2 \k_0 \le(\L_\b ~-~ \fr {\L_C}{\k}\ri); \nn\\
\la{lambdabetagammapr}
\L_{\c'} &=& \L_\c ~+~ 2 \k_0 \le(\L_\b ~-~ \fr {\L_C}{\k}\ri)
\er
and
\br
\la{kappa's}
\k_0 = \fr{c_0^2}{\kt}&;& \k_1 = \fr{c_1^2}{\kt} ~;~
\k_2 = \fr{c_0 c_1}{\kt} ~,~ \kt = c_0^2 + c_1^2 \k \, .~~
\er
$\L_{\b'}$ and $\L_{\c'}$ are $(N-n) \times (N-n)$ matrices, and
constants $(\k_0, \k_1, \k_2)$ describe the amount of mixing between
the GS and ES.

With these definitions, the density matrix (\ref{gses-den1}) can be
rewritten as
\be \la{gses-den2}
\r (t; t') ~=~ \kt ~ F (t; t') ~ \r_0 (t; t') ~.
\ee
As for the ES, here too the pre-factor $F (t; t')$ of the Gaussian $\r_0 
(t; t')$ cannot be factorized into $(N-n)$ two HO density matrices. However, 
as discussed in Appendix \ref{ent-es}, if the vector $t^T$ is outside the 
maximum $t_{max}^T$, given by Eq. (\ref{tmax}) corresponding to the $3 
\sigma$ limits, the argument of $\r_0 (t; t')$ is negligible. Therefore, 
if the conditions (\ref{smallness}) as well as the conditions
\bea \la{smallness1}
\et_1 \equiv t_{max}^T  \L_{\b'}  t_{max}  \ll 1~,~ 
\et_2 \equiv t_{max}^T  \L_{\c'}  t_{max}~ \ll 1 
\eea
are satisfied, then we can approximate the pre-factor $F (t; t')$ as 
\be \la{F-approx}
F (t; t') ~\approx~ \exp \le[\kt_1 ~w (t; t') ~+~ \kt_2 ~v (t; t')\ri],
\ee
where we have kept terms up to quadratic order in $t, t'$. [Note that,
$v (t; t')$ is only linear in $t, t'$ whereas $w (t; t')$ is quadratic
in $t, t'$.]

Using Eq. (\ref{gs-den}) for $\r_0 (t; t')$ we can now write the 
(approximated) MS density matrix as
\be \la{gses-den3}
\r (t; t') ~=~ \kt \sq{\fr{|\O|}{\pi^{N-n} |A|}} ~
\exp \le[z (t; t') + \kt_2 v (t; t')\ri]
\ee 
where
\br 
\la{z}
 \hspace*{-0.6cm}
z (t; t')= - ~ \fr{t^T \c' t + t'^T \c' t'} 2 ~+~ t^T \b' t' 
\er
and
\br
\la{betagammapr}
\hspace*{-0.6cm} 
\b' &=& \b + \kt_1 \L_{\b'} =\b + \kt_1 \L_\b - 2 \kt_0 \kt_1 
\le(\L_\b - \frac{\L_C}{\k}\ri) \nn\\
\hspace*{-0.6cm}
\c' &=& \c + \kt_1 \L_{\c'} = \c + \kt_1 \L_\c + 2 \kt_0 \kt_1 
\le(\L_\b - \frac{\L_C}{\k}\ri) 
\er
are $(N-n) \times (N-n)$ matrices.  $\b'$ is symmetric while $\c'$ 
is not necessarily symmetric.

Let us make the following transformations on the set of $(N-n)$
variables $t \equiv \{x_{n+1}, \cdots, x_N\}$ and $t' \equiv
\{x'_{n+1}, \cdots, x'_N\}$:
\be \la{var-shift}
t ~\rightarrow~ t ~+~ s ~;~~~ t' ~\rightarrow~ t' ~+~ s
\ee
where $s \equiv \{s_1,\cdots, s_{N-n}\}$ is a set of $(N-n)$ constant
values. The density matrix (\ref{gses-den3}) reduces to
\be \la{gses-den4}
\r (t; t') ~=~ \cN ~\exp \le[- ~ \fr{t^T \c' t + t'^T \c' t'} 2 ~+~ 
t^T \b' t' \ri]
\ee
where the normalization constant $\cN$ is given by
\be \la{new-norm}
\cN = \kt \sq{\fr{|\O|}{\pi^{N-n} |A|}} ~\exp 
\le[- s^T \le(\b' - \c'\ri)^T s\ri].
\ee
The $(N-n)$-dimensional constant column vector $s$ is determined 
from the equation
\be \la{s-eq}
s^T \le(\b' - \frac{\c' + \c'^T} 2\ri) = - \kt_2 \le(y_B - 
B^T A^{-1} y_A\ri).
\ee
It is easy to check that for either $c_0 = 0$ or $c_1 = 0$, the
constant $k_2$ vanishes, whence $s = 0$, and the density matrix
(\ref{gses-den4}) reduces either to that of pure GS \cite{sred} (for
$c_0 = 1, c_1 =0$, whence $\b' = \b, \c' = \c$) or that of ES
\cite{sdshankiES} (for $c_0 = 0, c_1 = 1$, whence $\b' = \b + \L_\b,
\c' = \c + \L_\c$). In general, when both $c_0$ and $c_1$ are
non-vanishing, then under the shifts $\b \rightarrow \b', \c
\rightarrow \c'$ (where $\b'$ and $\c'$ are given by
Eqs. (\ref{betagammapr})) the MS density matrix (\ref{gses-den4}) is
of the same form as the GS density matrix (\ref{gs-den}), up to a
normalization factor given above.  Such a normalization constant does
not affect the entropy computation.  Therefore we can use the same
steps as for GS [Eqs. (\ref{diag}) -- (\ref{gs-ent}), with the
replacements $\b \rightarrow \b', \c \rightarrow \c'$] to calculate
the total MS entropy.

\begin{figure*}[!htb]
\begin{center}
\epsfxsize 5.5 in
\epsfysize 2.25 in
\epsfbox{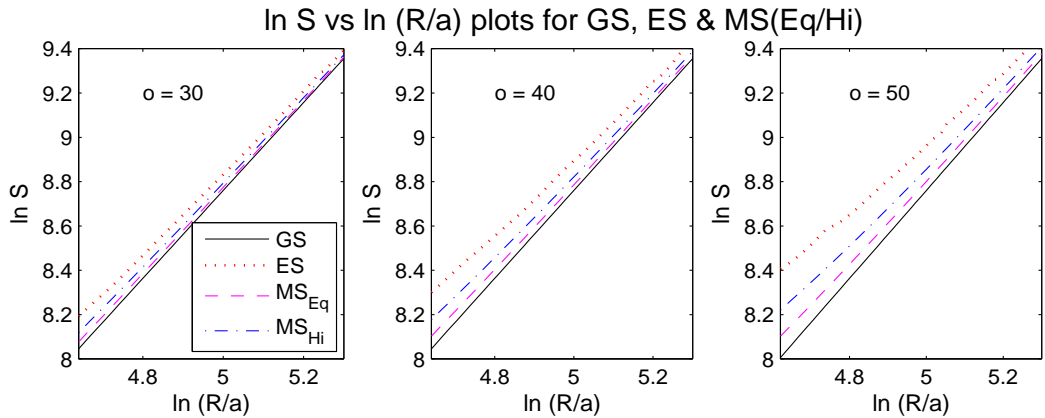}
\caption{Plots of logarithm of GS, ES and MS (Eq/Hi) entropies versus 
$\ln(R/a)$, where $R = a (n + 1/2)$ radius of the hypothetical sphere
(horizon), for $N = 300, ~ n = 100 - 200$ and $o = 30, 40, 50$ (in the
ES and MS cases). The numerical precision is $0.01\%$.}
\label{fig:1}
\end{center}
\vspace*{-0.05cm}
\end{figure*}

The rest of the analysis in this section is similar to that of Ref.
\cite{sdshankiES}. We compute the entanglement entropy numerically
(using MATLAB) in each of the cases: 
\begin{description}
\item (i) GS ($c_0 = 1, c_1 = 0$),
\item (ii) ES ($c_0 = 0, c_1 = 1$), 
\item (iii) an equal mixing (MS$_{Eq}$) of ES with GS ($c_0 = 
c_1 = 1/\sqrt{2}$), and 
\item (iv) a high mixing (MS$_{Hi}$) of ES with GS ($c_0 = 1/2, 
c_1 = \sqrt{3}/2$). 
\end{description}
The computations have been done with a precision\footnote{The
computations here are one order of magnitude more accurate than those
in Ref.\cite{sdshankiES}} of $0.01\%$ for the set of values: $N = 300,
~n = 100 - 200$ and $~o = 30, 40, 50$, $o$ being the last non-zero
columns of the vector $\a^T$. The conditions (\ref{smallness}) as well
as (\ref{smallness1}) are satisfied for these values of the
parameters.

The expectation value of energy, $\cE$, for MS can be expressed as
\be \la{MS-energy}
\cE = \cE_0 + \fr {c_1^2} o \sum_{i=N-o+1}^N k_{Di}^{1/2} ~;
\ee
where $\cE_0 = \fr 1 2 \sum_{i=1}^N k_{Di}^{1/2}$ is the (zero-point)
GS energy. Now the excess of energy over the zero point energy is
given by the second term in the above equation (\ref{MS-energy}). As
the value of $c_1$ is between $0$ and $1$ and since the $k_{Di}$'s are
in ascending order ($k_{Di} > k_{Dj}$ for $i > j$), the fractional change
in energy $(\cE - \cE_0)/\cE$ is at most about $\sim 5\%$ (corresponding
to the extreme situation $c_1 = 1$, i.e., ES), for $N = 300$ and $o \sim 50$.
Moreover, since there are $o$ number of terms in the sum in the second term of
Eq.(\ref{MS-energy}), the excitation energy $(\cE - \cE_0)$ is of the order
unity (in units of $1/a$, where $a$ is the lattice spacing). Therefore if
$a$ is chosen to be of the order of Planck length, then the above energy
is of the order of Planck energy. The mass of a semi-classical BH, on the
other hand, is much larger than the Planck mass. Hence, one may safely
neglect the back-reaction of the scalar field on the background geometry.

\begin{figure*}[!htb]
\begin{center}
\epsfxsize 5.5 in
\epsfysize 2.25 in
\epsfbox{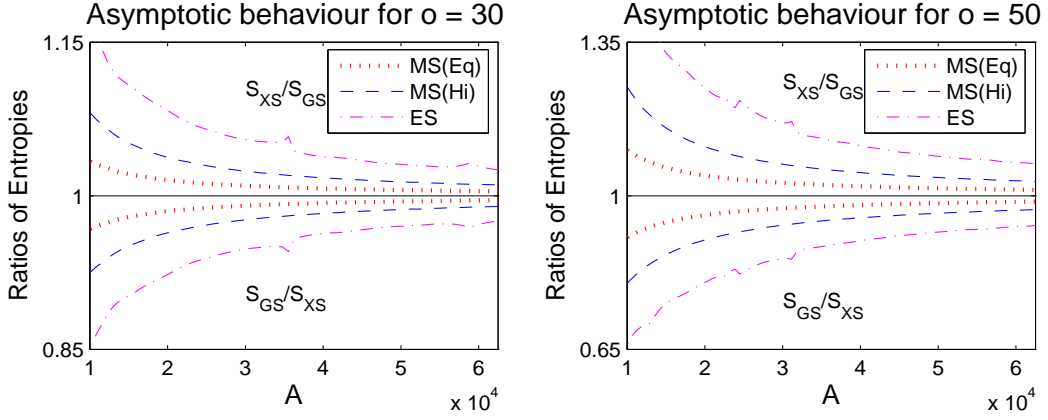}
\caption{Plots of ratios of GS and MS (Eq/Hi) or ES entropies and
their reciprocals versus the area $\cA$ (in units of $a^2$, $a$ being
the lattice spacing) for $o = 30, 50$. The plots show the asymptotic
nature of the MS and ES entropies with respect to the GS entropy. The
curves on the upper half (above $1$) show the variation of
$S_{XS}/S_{GS}$ with $\cA$, where XS stands for MS(Eq/Hi) or ES, while
the lower curves show the variation of $S_{GS}/S_{XS}$ with $\cA$.}
\label{fig:2}
\end{center}
\vspace*{-0.55cm}
\end{figure*}

In Fig. (\ref{fig:1}), we have plotted the logarithm of the entropy
$S$ versus $\ln(R/a) = \ln(n + 1/2)$, for different values of the
excitation ($o = 30, 40, 50$), for GS, ES and MS (Eq/Hi).  For GS, the
plot is very nearly the same as the numerical straight line fit
obtained in Ref. \cite{sred}, $S = 0.3 (R/a)^2$ with $N = 60$ lattice
points.  For the MS (Eq/Hi) cases, as well as for ES, the plots are
nearly linear for different values of the excitations $o = 30, 40, 50$
and appear to coincide with the plot for GS for large areas ($\cA = 4
\pi R^2 \gg a^2$). Numerical straight line fittings of the logarithm
of the ES entropy, $S_{ES}$, with $\ln(R/a)$ shown in
\cite{sdshankiES} revealed that for smaller areas $S_{ES} \sim
A^{\mu}$, where $\mu$ is always $< 1$ and decreases as the number of
excitation $o$ increases. To have a closer look at the behaviour of MS
entropy, $S_{MS}$, (for both equal and high mixings) and the ES
entropy $S_{ES}$ with respect to the GS entropy, $S_{GS}$, we have
plotted in Fig. (\ref{fig:2}) the ratios $S_{MS}/S_{GS},
S_{ES}/S_{GS}$ and the inverse ratios $S_{GS}/S_{MS}, S_{GS}/S_{ES}$,
versus the area $\cA$. For the range of excitations ($o = 30, 50$),
all the ratios tend to unity as the area increases. Thus the general
criterion of `asymptotic equivalence' \cite{asymp} is fulfilled, i.e.,

\be \la{asymp-eq}
\mbox{lim}_{_{\cA \rightarrow \infty}} \fr {S_{XS} (\cA)}
{S_{GS} (\cA)} = 1 ~;~ 
\mbox{lim}_{_{\cA \rightarrow \infty}} \fr {S_{GS} (\cA)}
{S_{XS} (\cA)} = 1~
\ee
where XS $\equiv$ MS (Eq or Hi) or ES. In other words, the MS (Eq/Hi)
and the ES entropies coincide asymptotically with the GS
entropy. However, as is evident from Fig. (\ref{fig:2}), the MS(Eq)
entropy is closer to the GS entropy for large $\cA$, than
the MS(Hi) entropy and the ES entropy, the latter being the
farthest. This implies that the asymptotic behaviour is strongly
influenced by the relative weight $c_1$ of the mixing of ES with GS
--- the smaller the value of $c_1$ the sharper is the asymptote.

%
%
\begin{figure*}[!htb]
\begin{center}
\epsfxsize 5.5 in
\epsfysize 2.25 in
\epsfbox{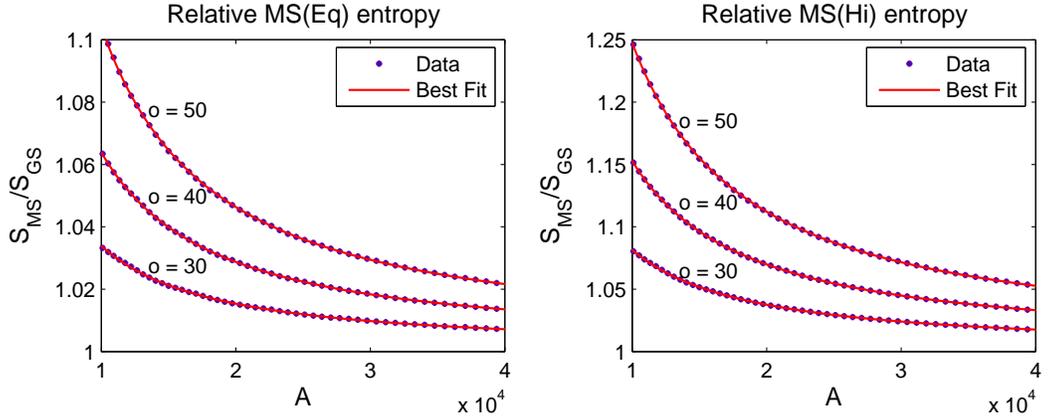}
\caption{Best fit plots (solid lines) of the relative mixed state entropies 
($S_{MS}/S_{GS}$) for equal and high mixings versus the area $\cA$ (in units of 
$a^2$), for $o = 30, 40, 50$. The corresponding data are shown by asterisks.}
\label{fig:3}
\end{center}
\vspace*{-0.55cm}
\end{figure*}
%
\begin{table*}[!htb]
\la{table1}
\begin{tabular}{|c||c|c|c||c|c|c|}
\hline
Fitting&\multicolumn{3}{c|}{For MS$_{Eq}$}&\multicolumn{3}{c|}{For MS$_{Hi}$} \\
\cline{2-7}
Parameters&$~o = 30~$&$~o = 40~$&$~o = 50~$&$~o = 30~$&$~o = 40~$&$~o = 50~$ \\
\hline\hline 
& & & & & &  \\
$\s_0$&$~1.001~$&$~1.002~$&$~1.003~$&$~1.001~$&$~1.004~$&$~1.006~$ \\
& & & & & &  \\
$\s$&$~1738~$&$~4288~$&$~8039~$&$~2956~$&$~7652~$&$~14120~$ \\  
& & & & & &  \\
$\n$&$~1.180~$&$~1.210~$&$~1.225~$&$~1.141~$&$~1.178~$&$~1.192~$ \\
\hline
\end{tabular}
\caption{Values of the parameters of the fit $S_{MS}/S_{GS} = \s_0 + \s \le(\cA/a^2\ri)^{-\n} $ 
for both MS(Eq) and MS(Hi) cases with the amounts of excitation $o = 30, 40, 50$.}
\end{table*}

%
\begin{figure*}[!htb]
\begin{center}
\epsfxsize 4.75 in
\epsfysize 3.5 in
\epsfbox{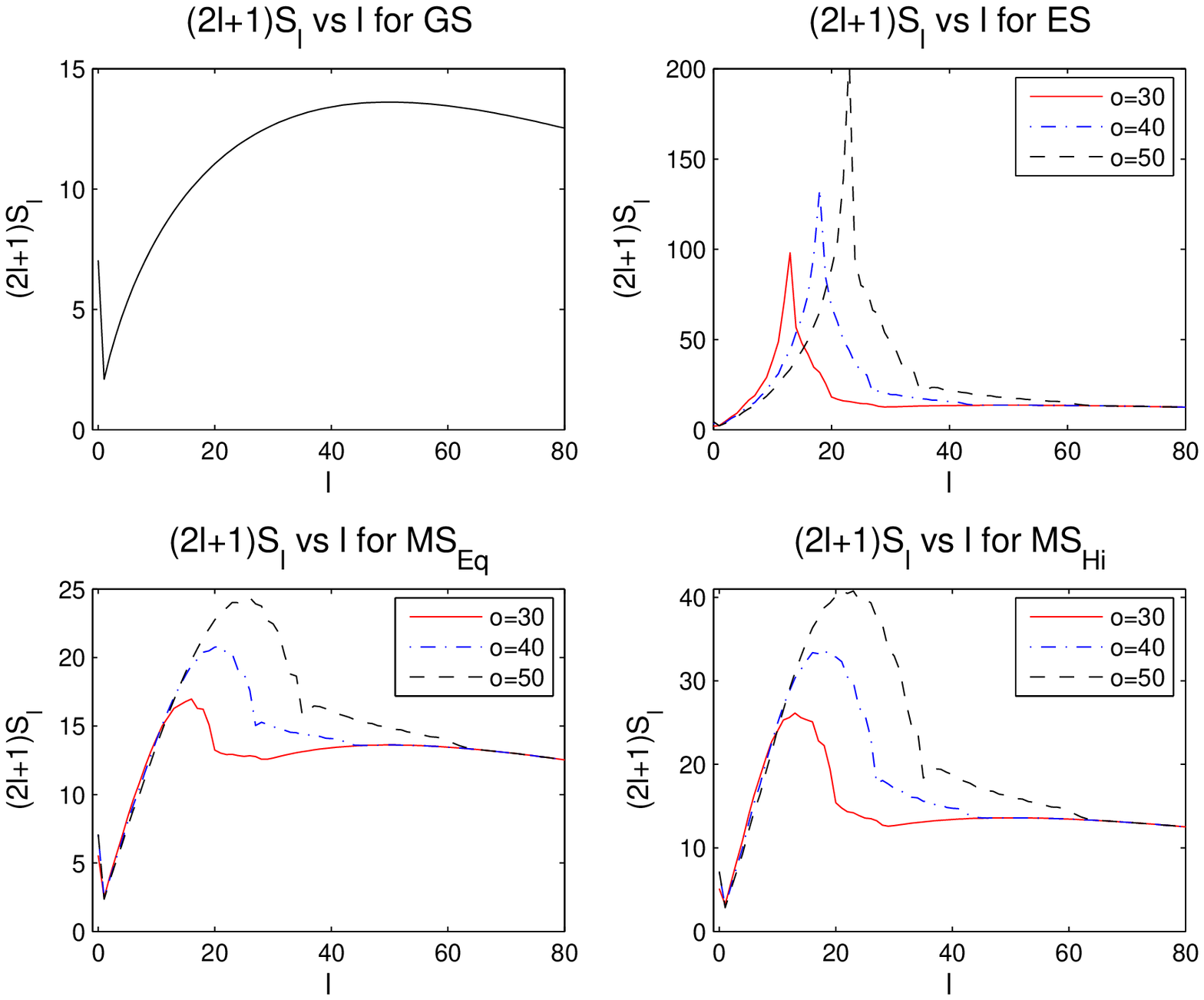}
\caption{Plot of the distribution of entropy per partial wave $[(2 1 +
    1) S_l]$ in the cases of GS, ES and MS (Eq/Hi), for $N = 300, n = 100$ 
    and $o = 30, 40, 50$.}
\label{fig:4}
\end{center}
\vspace*{-0.55cm}
\end{figure*}

In order to make things more transparent we have plotted in
Fig. (\ref{fig:3}) the best fit ratios of the MS entropies (for equal
and high mixings, with $o = 30, 40, 50$) to the GS entropy using a
simple formula:
\be \la{MS-fita}
\frac{S_{MS}}{S_{GS}} =  \s_0 ~+~ \s \le(\fr{\cA}{a^2}\ri)^{-\n} ~.
\ee
The fitting parameters $\s_0, \s$ and $\n$ are shown in Table 1. The
parameter $\s_0$ is very close to unity for all values of $o = 30, 40,
50$, for both MS(Eq) and MS(Hi) cases.  However, there is a slight
increase in $\s_0$ as $o$ increases or for greater relative weight
$c_1$ of mixing of ES with GS for a particular $o$ (i.e., $\s_0$ is
greater in the MS(Hi) case than in the MS(Eq) case for fixed
$o$). Neglecting this variation in $\s_0$ and noting that the GS
entropy can be written as $S_{GS} = n_0 (\cA/a^2)$, where $n_0$ is a
constant, we can approximately express:
\be \la{MS-fit}
S_{MS} ~=~ S_{GS} + \tilde{\s} \le(\fr{\cA}{a^2}\ri)^{1-\n}~,
\ee
where $\tilde{\s} = n_0 \s$. As the value of exponent $(1 - \n)$ lies
between $0$ and $-1$ for both equal and high mixings (see Table 1) the
second term in the above Eq. (\ref{MS-fit}) may be regarded as a {\it
power law correction} to the AL, resulting from entanglement, when the
wave-function of the field is chosen to be a superposition of GS and
ES. It is important to note that the correction term falls off rapidly
with $\cA$ (due to the negative exponent) and in the semi-classical
limit ($\cA \gg 1$) the AL is hence recovered. This lends further
credence to entanglement as a possible source of black-hole
entropy. The correction term is more significant for higher
excitations $o$ or greater ES-GS mixing proportion $c_1$. This is
evident from Table 1, which shows that the parameter $\s$ (and hence
$\tilde{\s}$) increases and the parameter $\n$ (and hence the negative
exponent $|1 - \n|$) decreases with the increase in $o$ (fixed $c_1$)
or the increase in $c_1$ (fixed $o$).

Fig. (\ref{fig:4}) shows the variation of $(2l+1)S_l$ with $l$, in the
cases of GS, ES and MS (Eq/Hi) for a fixed $n (= 100)$ and a set of
increasing values of $o$. For the GS, there is a peak at $l=0$
($s$-wave), followed by another one at $l \approx 40$ due to the
degeneracy factor $(2l+1)$. The first peak shifts to a value $l > 0$
for the ES, and the shift is greater as $o$ is increased. There is,
however, no second peak in this case, although there seems to be an
increase towards higher values of $l$. Thus, higher partial waves are
seen to get excited with greater excitations. In each of the MS cases,
there is a trace of the first peak at $l = 0$ as for GS, however the
amplitude of that peak is very small compared to the second peak which
appears between $l \sim 10 - 30$ depending on the value of the
excitations $o$. As in the case of ES, the second peaks for MS (Eq/Hi)
are higher and far away from $l = 0$ for increasing values of
$o$. However, relative to the ES case, there is a broadening of the
half-width of the peaks for MS, though not as broad as that of the
second peak for GS. Thus, as expected, the $(2l+1)S_l$ vs $l$ curves
for MS show features that are intermediate between those for GS and
ES.

\section{Location of the degrees of freedom}
\la{dof}

\begin{figure*}[!htb]
\begin{center}
\epsfxsize 4.75 in
\epsfysize 3.5 in
\epsfbox{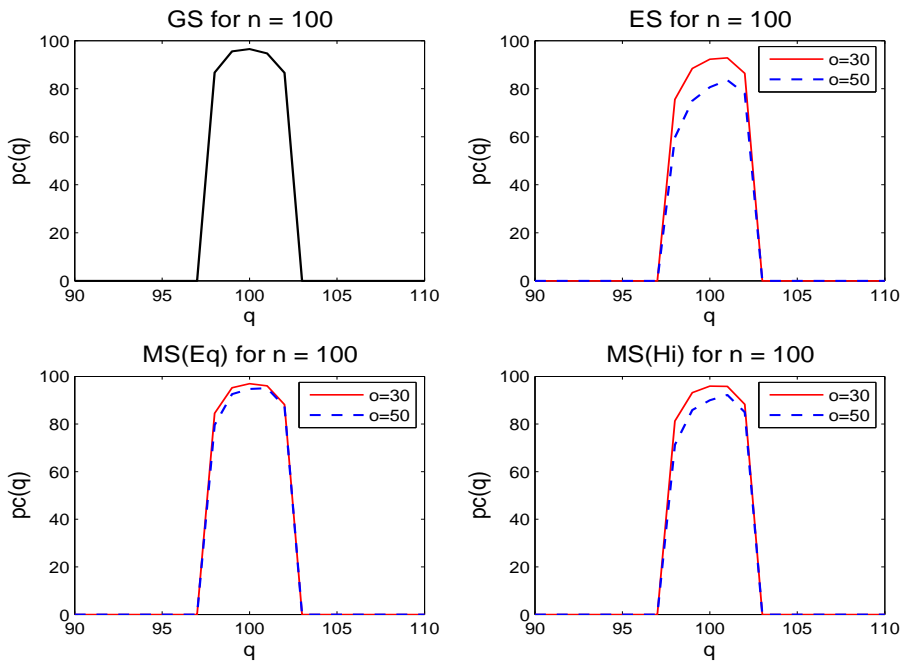}
\caption{Plots of the percentage contribution $pc(q)$ to the total entropy 
as a function of window position $q$, for a window size $d = 5$ and
fixed $N = 300, n = 100$, in each of cases of GS, ES and MS
(Eq/Hi). For ES and MS (Eq/Hi) the solid curve is for $o = 30$ whereas
the broken curve is for $o = 50$.}
\label{fig:5}
\end{center}
\vspace*{-0.55cm}
\end{figure*}

Let us now examine closely the expression for the interaction matrix
$K_{ij}$, Eq. (\ref{kij}), for the system of $N$ HOs. The last two terms
which signify the nearest-neighbour (NN) interaction between the
oscillators, are solely responsible for the entanglement entropy of
black-holes, i.e., $S_{BH} = 0$ if these two terms are set to
zero. Let us, however, consider the situation where the NN
interactions, and hence the off-diagonal elements of $K_{ij}$, are set
to zero (by hand) everywhere except in a `window', such that the
indices $i,j$ runs from $q - s$ to $q + s$, where $s \leq q$. Thus the
interaction region is restricted to a width of $d = 2s + 1$ radial
lattice points. Now, choosing the position of the center of the window
$q$ to vary between $0$ and a value $q_{max} > n$, we allow the window
to move rigidly across from the origin to a point outside the
horizon. Fig. (\ref{fig:5}) shows the variation of the percentage
contribution of the entropy for a fixed window size of $5$ lattice
points ($d = 5, s = 2$), i.e.,
\be
pc (q) ~=~ \fr{S (q, d=5)}{S_{tot}} \times 100 
\ee
as a function of $q$ for fixed values $N = 300, n = 100$ in each of
the cases GS and ES, MS (Eq/Hi) with $o = 30, 50$. Here $S_{tot}$ is
the total entropy with all the NN interactions present, i.e., $i,j$
running from $0$ to $N$.

%

In all the cases of GS, ES and MS (Eq/Hi) the first observation is
that $pc (q) = 0$ when $q$ is far away from $n$. 
There is no contribution to the total
entanglement entropy if interaction window does not include the
horizon. For values of $q$ very close to $n$ there are significant
contributions to $S_{tot}$ and in the case of GS, $pc (q)$ peaks
exactly at $q = n$. For ES and MS, however, 
the peaks tends to shift towards a value $q > n$, its exact position 
depending on the amount of excitation $o$. Moreover, the amplitudes of
the peaks gradually diminish as the value of $o$ and/or the mixing
weight $c_1$ increases. Thus, we infer that:
\bi
\item The contribution to the total entropy is more from the DoF that are 
in the vicinity (inside or outside) of the horizon, rather than far from it.
\item The contributions, however small, from the DoF far away from the horizon 
are more for MS(Eq/Hi) and ES, compared to the GS.
In other words, the contributions from the far away DoF increases with 
increase in the number of excitations and amount of mixing of ES with GS.
\ei

Further investigations have been carried out recently in Ref. 
\cite{sdshankiDoF} to check the effects of the far-away DoF on the
total entropy, by keeping fixed the center of the window at the horizon,
i.e., $q = n$, while increasing the window width $d$ from $0$ to $n$. It
is found that for GS about $85\%$ of the total entropy is obtained within 
a width of just one lattice spacing, and within a width of $d = 3$ 
almost the entire GS entropy is recovered. Thus most of the GS entropy 
comes from the DoF very close to the horizon and a small part (about 
$15\%$), has its origin deeper inside. For ES, however, the 
corresponding figures are about $60\%$ ($d=5$), 
and the total ES entropy is recovered when $d$ is as much as $15 - 20$,
depending on the number of excitations $o = 30 - 50$. Thus the far-away
DoF contribute more to the entropy for the ES.
This, in turn, may be
looked upon as follows: {\it larger the deviation from the area
law, larger is the contribution to the total entropy from the
DoF that are far away from the horizon}.
The situation is intermediate for the MS 
(which itself interpolates between the GS and ES): 
This is evident
from Fig. (\ref{fig:5}) [and also from Fig. (\ref{fig:1})] where unlike
the curves for ES, those for the MS cases do not show much deviations
from the curve for GS, even for high excitations $o$.

\section{Entanglement entropy of massive scalar field}  
\la{ent-massive}
      
As shown in Appendix \ref{app:sf-mot}, the equation of motion for
metric perturbations in a general space-time with a cosmological
constant $|\Lambda|$ coincides with that of a test massive scalar field 
propagating in the background metric. In all our earlier analysis, we 
had set, for simplicity, $|\Lambda| = 0$. In this section, we obtain the 
entanglement entropy for the massive scalar field.

%
\begin{figure*}[!htb]
\begin{center}
\epsfxsize 4.75 in
\epsfysize 3.5 in
\epsfbox{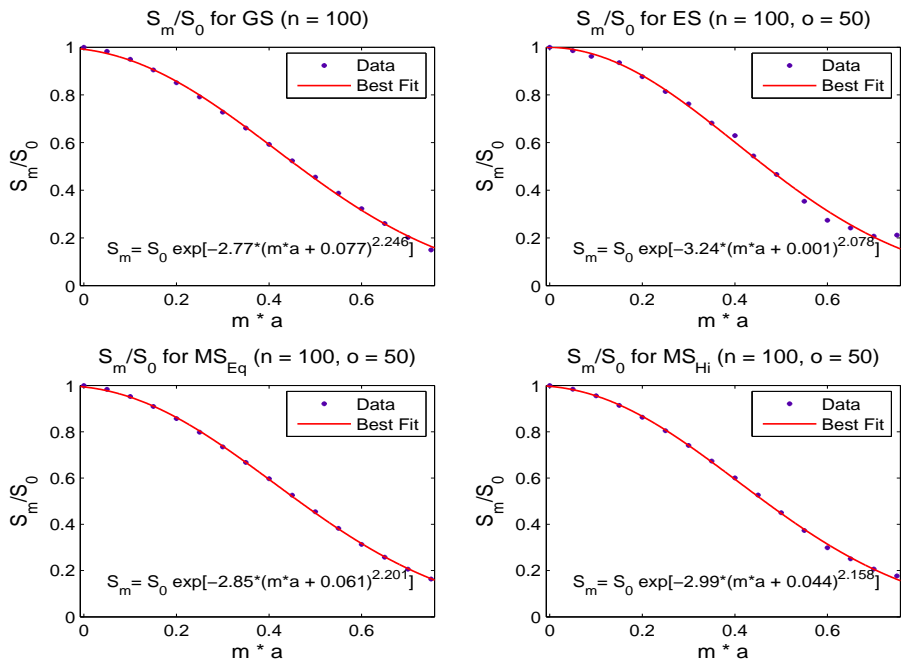}
\caption{Best fit plots of the relative variation of the total entropy 
$S_m$ for a massive scalar field (in units of the total entropy $S_0$
corresponding to a massless scalar field) with the mass $m$ times the 
lattice spacing $a$, for fixed $n = 100, o = 30$, in each of cases of 
GS, ES and MS (Eq/Hi). The corresponding data are shown by asterisks. 
The fits show an exponential damping of the ratio $S_m/S_0$ with mass.}
\label{fig:6}
\end{center}
\vspace*{-0.55cm}
\end{figure*}

The action for the massive scalar field (mass $m$) propagating in
the background space-time $g_{\mu\nu}$ is
\beq
S = -\frac{1}{2}
\int d^4 x \, \sqrt{-g}~ \le[g^{\mu\nu}~ \pa_{\mu}\vph~\pa_{\nu} \vph 
+ m^2 \vph^2 \ri]
\la{act-mass}
\eeq
Repeating the steps described in Appendix \ref{app:BH-Ham} for the
massive scalar will lead to massive, free field Hamiltonian
(\ref{ham2}). Discretizing the resulting Hamiltonian, as described in
Sec. (\ref{ent-scalar}) will lead to the $N-$coupled HO Hamiltonian,
with the interaction matrix $K_{ij}^{(m)}$ given by:
\bea \la{kij-mass}
K_{ij}^{(m)} ~=~ K_{ij} ~+~ \le(m a\ri)^2
\eea
where $K_{ij}$ is the interaction matrix, given by Eq.(\ref{kij}), for 
massless scalar field.
Following the steps discussed in sections (\ref{ent-scalar}) and
(\ref{ent-gses}), we can compute the entanglement entropy for the
massive field ($S_{m}$).

In Fig. (\ref{fig:6}), we have plotted $S_m/S_0$ [where, $S_0$ is
the entropy for the massless scalar] for the GS, ES and MS(Eq/Hi) for
$N = 300, n = 100, o = 50$. The Gaussian fits indicate an exponential
fall-off of $S_m$ with respect to $S_0$ as the mass increases:
\be
S_m ~=~ S_0 ~\exp\le[- \a_1 \le(m a ~+~ \a_2\ri)^\l\ri]
\ee
where $\a_1, \a_2$ and $\l$ are the fitting parameters. Depending on
the state (GS, ES or MS), the parameter $\a_1$ varies between $2.77$
and $3.24$, $\a_2$ is between $0.077$ and $0.001$ and the power $\l$ 
is close to $2$. Thus approximately $S_m/S_0$ scales as $e^{- m^2 a^2}$.
There is, however, a small variation in the power $\l$ for the different
cases. The exponential damping is strongest for GS, and gradually
slows down as more and more ES oscillators are mixed with GS, the
damping is slowest for the ES case.

Although $S_m$ scales as $S_0$ times a mass-dependent exponential
term, the fitting parameters $\a_1, \a_2$ and $\l$ change very little
for the different cases GS, MS(Eq/Hi) and ES, even for a fairly high
amount of excitation $o = 50$. As such, for a fixed mass $m$, the
variations $\ln S$ versus $\ln (R/a)$ for all the cases remain almost
the same as those for massless scalar field, cf. Fig. (\ref{fig:1}). 
The analysis and inferences of the previous sections for massless scalar 
go through for the massive scalar field, resulting in correction terms 
obtained before.

\section{Conclusions}       
\la{conclu}

In this work, we have obtained power-law corrections to entanglement
entropy, which may be relevant for the entropy of BH. Indeed, as shown 
in Appendices \ref{app:sf-mot} and \ref{app:BH-Ham}, certain modes of 
gravitational perturbations in black-hole space-times behave as minimally 
coupled scalar fields. Also for computations of time-independent quantities 
done at a fixed value of Lema\^itre time, it suffices to consider an 
effective flat space Hamiltonian. Extending the analysis of the earlier 
work \cite{sdshankiES}, we have shown that for small black-hole areas the 
area law is violated not only when the oscillator modes that represent the
scalar DoF are in ES, but also when they are in a linear superposition of 
GS and ES. We found that the corrections to the AL become increasingly 
significant as the proportion of ES in the superposed state increases.  
Conversely, for large horizon areas, these corrections are relatively small 
and the AL is recovered.

It is interesting to compare and contrast the power-law corrections
obtained here to those derived in the case of higher-derivative gravity
\cite{Wald:1993a}. The power-law corrections to the Bekenstein-Hawking
entropy derived in both --- entanglement and higher derivative gravity
-- these approaches have same features. For instance, it was shown
that the entropy of 5-dimensional Boulware-Deser black-hole
\cite{HDentropy} is given by
\begin{eqnarray}
S = \frac{A}{4} \le[1 + \frac{c}{A^{2/3}} \ri] \qquad ;~~~~
c = \mbox{~constant} \, .
\end{eqnarray}
As in Eq. (\ref{MS-fit}) the above entropy is proportional to area for
large horizon radius, however it strongly deviates in the small
horizon limit. It is important to note that the corrections to the
black-hole entropy are generic and valid even for black-holes in
General relativity without any higher curvature terms\footnote{In this
context, it should be mentioned that it is not possible to check for
logarithmic corrections to the entropy in our analysis, as the
numerical error we obtain is much larger than $\ln (n + 1/2)$.}.
It is interesting to investigate the relation between the entanglement
entropy with that of Noether charge approach \cite{sdshankisscurv}.

The location of the DoF that give rise to the entanglement entropy has
important implications as far as the corrections to the AL is
concerned.  It is found that for GS, ES and MS, the contributions to
the total entropy from the DoF that are nearest to the horizon are
maximum. However, there are small contributions from DoF that are far
away from the horizon, which also need to be taken into account in
order for the AL to emerge. These contributions are least in the case
of GS and gradually increase as the proportion of mixing of ES with GS
and/or the amount of excitation increases. Correspondingly, there are
increasing deviations from the AL. Thus one is led to conclude that
the AL is intimately linked with near horizon DoF.

We have also shown that the mass of the scalar field
does not have much influence on the corrections to the AL. The total
entropy for the massive field scales as that of the massless field
times a mass-dependent exponentially damping term that varies very
slowly with the mixing proportion and the amount of excitation which
are key to produce the AL corrections.

Open problems in the context of entanglement entropy include: (i) The
proportionality constant in the relation $S = 0.3 (R/a)^2$ for GS
obtained in ref.\cite{sred} differs from the $1/4$ in the
Bekenstein-Hawking relation [Eq.(\ref{temp1})].  This discrepancy
persists for MS and ES.  A probable reason behind this mismatch is the
dependence of the pre-factor on the type of the discretization scheme.
For example, another discretization scheme, resulting in the NN
interactions between four or more immediate neighbours, would result
in a different pre-factor. Is it then at all possible to obtain the
Bekenstein-Hawking value?  (ii) Can a temperature emerge in the
entanglement entropy scenario, and if so, then along with the current
entropy, will it be consistent with the first law of BH
thermodynamics?  (iii) Are the second and third laws of thermodynamics
valid for this entropy?  (iv) Can the entanglement of scalar fields
help us to understand the evolution or dynamics of BHs and the
information loss problem?  We hope to report on these in future.

\bigskip
\acknowledgments
The works of SD and SSu are supported by the Natural Sciences and 
Engineering Research Council of Canada.

\bigskip
\noindent
{\bf Note added:} Recently, in Ref. \cite{Sarkar:2007uz}, the authors 
have calculated the sub-leading power-law corrections to the 
Bekenstein-Hawking entropy using the canonical ensemble (aka Brick-wall)
approach \cite{thooft}. The results reported there agree with the 
numerical results derived in this work.


\appendix

\section{Why consider scalar fields?}
\label{app:sf-mot}

In this appendix, we briefly discuss the motivation for obtaining the
entanglement entropy of a scalar field. First, we obtain the equation
of motion of the metric perturbations for a general space-time and
then, as a special case, discuss the equation of motion of the
perturbations in asymptotically flat spherically symmetric
space-times.

Consider the Einstein-Hilbert action with a positive cosmological
constant ($|\Lambda|$):
\beq
\label{eq:EHAction}
S_{_{EH}} (\bar{g}) = M_{_{\rm Pl}}^2 \int d^4x \sqrt{-\bar{g}}
\le[\bar{R} - 2 |\Lambda|\ri] \, .  \eeq
Let $\bar{g}_{\mu\nu} = g_{\mu\nu} + h_{\mu\nu}$ and expand the
action keeping only the parts quadratic in $h_{\mu\nu}$, we get
\cite{barthchristen}
\br
\label{eq:PerEH}
S_{_{EH}} (g, h) &=& - M_{_{\rm Pl}}^2 \! \int \!\! d^4x \sqrt{|g|}\, 
\le[ 2\, \gamma^\a_{~\mu\nu}\, \gamma_{~~\a}^{\mu\nu} \ri. \nn \\
& & \le. \quad + \frac{1}{4}\nabla_{\mu}\tilde{h}
 \nabla^{\mu}\tilde{h} + \frac{|\Lambda|}{2} h_{\mu\nu}
 \tilde{h}^{\mu\nu} \ri] \er
where
\br
\label{htilde}
\tilde{h}_{\mu\nu} &\equiv& h_{\mu\nu}-\frac{1}{2} g_{\mu\nu}h_{\a}^{\a} \, ,
\qquad \tilde{h} \equiv \tilde{h}^{\mu}_{\mu} \, ,\\
\gamma^\a_{~\mu\nu} &\equiv& \frac{1}{2}
(\nabla_{\mu}\tilde{h}_{\nu}^{\a}+\nabla_{\nu}\tilde{h}_{\mu}^{\a} 
- \nabla^{\a}\tilde{h}_{\mu\nu}) \, .
\er
The above action is invariant under the infinitesimal gauge transformation 
$h_{\mu\nu} \to h_{\mu\nu} + \nabla_{(\mu} \xi_{\nu)}$ when the background 
metric $g_{\mu\nu}$ satisfies the vacuum Einstein's equation with $|\Lambda|$. 
We can remove the gauge arbitrariness by imposing the harmonic gauge condition 
$\pa_{\mu} \tilde{h}^{\mu\nu} = 0$ \cite{barthchristen}.

Assuming $h_{\mu\nu}$ to be small, we can keep only the first
derivatives of $h_{\mu\nu}$. With these two conditions, the action
(\ref{eq:PerEH}) reduces to (for more details see Ref. \cite{lem},
pgs. 330-332)
\beq
S_{_{EH}}(g, h) = - \frac{M_{_{\rm Pl}}^2}{2} \!\! \int \!\!\! d^4x \sqrt{|g|}\, 
\le[\nabla_{\alpha} {h}_{\mu\nu} \nabla^{\alpha} {h}^{\mu\nu} 
+ |\Lambda| h_{\mu\nu} h^{\mu\nu}\ri]\!\! \, .
\eeq
The above action corresponds to the massive spin-2 $h_{\mu\nu}$ 
field propogating in the background metric $g_{\mu\nu}$ where the
cosmological constant appears as mass-term. In the weak field limit --- 
when the gravitational field is weak like in the case of regions close 
to the black-hole horizons --- ${h}^{\mu\nu}$ can be approximated as 
a plane-wave perturbation with a particular frequency, i. e. 
$h_{\mu\nu} = M_{_{\rm Pl}}^{-1} \epsilon_{\mu\nu} \varphi(x^{\mu})$ 
[where $\epsilon_{\mu\nu}$ is the constant polarization tensor], the 
above action can be written as 
\beq
S_{_{EH}} (g, h) = - \frac 1 2 \int \!\!\! d^4x \sqrt{|g|}\, 
\le[\pa_{\alpha} \vph \pa^{\alpha} \vph 
+ |\Lambda| \vph^2 \ri] \, .
\eeq
which is the action for the massive scalar field propagating in the
background metric $g_{\mu\nu}$. 

Moreover, for four-dimensional spherically symmetric space-times,
the metric perturbations are of two kinds --- axial and polar
\cite{chandra,qnm,perturb}. The equations of motion of both these
perturbations are scalar in nature and are related to each other
by a unitary transformation \cite{chandra}. The equation of motion
of the axial perturbations is nothing but the equation of motion
of a test, massless scalar field propagating in the black-hole
background:
\be
\Box \vph \equiv \fr 1 {\sq{-g}} \pa_\m 
\le(\sq{- {g}} {g}^{\m\n} \pa_\n 
\vph\ri) = 0.
\ee

Hence, by computing the entanglement entropy of the scalar fields,
we obtain the entropy of a class of metric perturbations of the
background space-time. Of course, a generic perturbation being a 
superposition of plane wave modes, and entanglement entropy being 
a non-linear function of the wave-function, we do not claim that 
such a computation would account for the entropy of all perturbations. 
Nevertheless, it is expected to shed important light on the role of 
entanglement in the AL.

In most part of this work, we calculate the entropy of the massless
scalar field (i.e., setting $|\Lambda| = 0$). In
Sec. (\ref{ent-massive}), we obtain the entanglement entropy of the
massive field, corresponding to $|\Lambda| \neq 0$.

\section{Hamiltonian of scalar fields in black-hole space-times}
\la{app:BH-Ham}

In this appendix, we obtain the Hamiltonian of the massless scalar
field propagating in a general static spherically symmetric
space-time. We show that for a particular choice of time slicing 
such a Hamiltonian reduces to the Hamiltonian of a scalar field in 
flat space-time. 

In Ref. \cite{sdshankiDoF}, two of the authors (SD and SSh) showed
that in a fixed Lema\^itre time coordinate, the Hamiltonian of the
scalar field propagating in Schwarzschild space-time reduces to
the scalar field Hamiltonian in flat space-time. In this appendix,
we extend the analysis for any non-degenerate static spherically
symmetric space-times.

Let us consider the following line-element:
\bea \la{bh-metric}
ds^2 &=& - A(\tau,\xi) \, d\tau^2 + 
\fr{d\xi^2}{B(\tau,\xi)} + \r^2(\tau,\xi) d\O^2  
\eea   
where $A, B, \r$ are continuous, differentiable functions of
$(\tau,\xi)$ and $d\Omega^2 = d\theta^2 + \sin^2 \theta d\phi^2$ 
is the metric on the unit $2-$sphere. The action for the scalar 
field propagating in the above background is given by
\br
S &=& -\frac{1}{2}
\int d^4 x \, \sqrt{-g}~ g^{\mu\nu}~ \pa_{\mu}\vph~\pa_{\nu}\vph \nn \\
\label{eq:actgen1}
&=& - \frac{1}{2} \sum_{l m} 
\int d\tau d\xi \Bigl[ -\frac{\rho^{2}}{\sqrt{A \, B}} 
(\pa_{\tau}\vph_{_{lm}})^2 \Bigr. \\
& & + \Bigl. \sqrt{A B} \rho^{2} ~(\pa_{\xi}\vph_{_{lm}})^2 
+~ l(l + 1) \sqrt{\frac{A}{B}} \,
\vph_{_{lm}}^2 \Bigr] \, . \nn
\er
where we have decomposed $\vph$ in terms of the real spherical
harmonics ($Z_{lm}(\th, \f)$):
\be \la{tens-sph}
\vph (x^{\mu}) = \sum_{l m} \vph_{_{lm}}(\tau,\xi) Z_{l m} (\th, \f) \, . 
\ee
Following the standard rules, the canonical momenta and Hamiltonian of
the field are given by
\br
\la{eq:mom}
{\Pi}_{_{lm}}&=& \frac{\pa \cal{L}}{\pa(\pa_{\tau} \vph_{lm})} =
\frac{\rho^{2}}{\sqrt{A \, B}} \, \pa_{\tau} \vph_{_{lm}} \, ,\\
\la{eq:gen-Ham}
H_{lm}(\tau) &=& \!\! \frac{1}{2} \int_{\tau}^{\infty} \!\!\!\!\!\!
d\xi \! \le[\! \frac{\sqrt{A B}}{\rho^{2}} \Pi_{_{lm}}^2 
+ \sqrt{A B} \, \rho^{2} (\pa_{\xi} \vph_{_{lm}})^2 \ri. \\
& +& \le.l(l + 1) \sqrt{\frac{A}{B}} \, \vph_{_{lm}}^2 \ri] \, ,
\qquad H = \sum_{lm} H_{lm} \, . \nn
\er
The canonical variables $(\vph_{_{lm}}, \Pi_{_{lm}})$ satisfy the
Poisson brackets
\br
\label{eq:gen-PB}
& & \{\vph_{_{lm}}(\tau,\xi), \Pi_{_{lm}}(\tau,\xi')\} = \delta(\xi - \xi') \\
& & \{\vph_{_{lm}}(\tau,\xi), \vph_{_{lm}}(\tau,\xi')\} = 0 = 
\{\Pi_{_{lm}}(\tau,\xi), \Pi_{_{lm}}(\tau,\xi')\} \, .\nn
\er
Having obtained the general Hamiltonian, our next step is to show that
this reduces to the flat space-time Hamiltonian of the scalar field in
a fixed Lema\^itre time. In the time-dependent Lema\^itre coordinates 
\cite{lem,shanki:2k3} the line-element is given by (\ref{bh-metric}) 
with
\beq
A(\tau,\xi) = 1;~ B(\tau,\xi) = \frac{1}{1 - f(r)};~
\rho(\tau,\xi) = r \, ,
\eel{eq:lemcoor}
where $r = r(\tau,\xi)$.

The line-element in Lema\^itre coordinates is related to that in the 
time-independent Schwarzschild coordinates, viz.,
\be \la{sch1}
ds^2 = - f(r) dt^2 + \fr{dr}{f(r)} + r^2 d\O^2 \,;  \quad  f(r = r_h) = 0
\ee
by the following transformation relations \cite{shanki:2k3}:
\br
\tau = t \pm \int\!\! dr \frac{\sqrt{1 - f(r)}}{f(r)}&;& 
\xi = t + \int\!\! dr \frac{[1 - f(r)]^{-1/2}}{f(r)} \nn \\
\label{eq:xitau}
\xi - \tau &=& \int \frac{dr}{\sqrt{1 - f(r)}}
\er
The advantage of the Lema\^itre coordinate over the Schwarzschild
coordinate is that (i) the former is not singular at the horizon $r_h$ 
as opposed to the latter, and (ii) $\xi$(or, $\tau$) are space(or, time)-like 
everywhere while $r$(or, $t$) is space(or, time)-like only for $r > r_h$.  

Substituting the relations (\ref{eq:lemcoor}) in the general
Hamiltonian (\ref{eq:gen-Ham}), we get,
\br
\la{eq:lem-Ham}
H_{_{lm}}(\tau) &=& \frac{1}{2} \int_{\tau}^{\infty} \!\!\!\!
d\xi \le[ \frac{1}{r^2 \sqrt{1 - f(r)}} \Pi_{_{lm}}^2  \ri. \\
&+& \!\!\!\! \le. \frac{r^{2}}{\sqrt{1 - f(r)}} \le(\pa_{\xi} \vph_{_{lm}}\ri)^2 
 + l(l + 1)\sqrt{1 - f(r)} \, \vph_{_{lm}}^2 \ri] \nn
\er
where the conjugate variables satisfy the Poisson brackets
(\ref{eq:gen-PB}). Note that the scalar field and the Hamiltonian
depend explicitly on the Lema\^itre time. 

Next, choosing the the Lema\^itre time ($\tau = \tau_0 = 0$), the
relations (\ref{eq:xitau}) lead to:
\beq \la{dxi}
\frac{d\xi}{dr} = \frac{1}{\sqrt{1 - f(r)}} \, .
\eeq
If we set $d\theta = d\phi = 0$, then for the fixed Lema\^itre time $\tau_0$
it follows that $ds^2 = d\xi^2/B(\tau_0,\xi) = dr^2$, i.e., the covariant cut-off 
is $|ds| = dr$.
 
Substituting the above relation (\ref{dxi}) in the Hamiltonian
(\ref{eq:lem-Ham})
we get
\beq
H_{_{lm}}(0) = \frac{1}{2} \int_{0}^{\infty} \!\!\!\!\! dr \!\!
\le[\frac{\Pi_{_{lm}}^2 r^{-2}}{1 - f(r)} + r^2 \le(\pa_r \vph_{_{lm}}\ri)^2 
+ l(l + 1) \, \vph_{_{lm}}^2\ri]
\eeq
where the variables $(\vph_{_{lm}}, \Pi_{_{lm}})$ satisfy the relation:
\beq
\{\vph_{_{lm}}(r), \Pi_{_{lm}}(r')\} = \sqrt{1 - f(r)} \delta(r - r') .
\eeq

Performing the following canonical transformations
\beq
\Pi_{_{lm}} \to {r \sqrt{1 - f(r)}} \, \Pi_{_{lm}}  \, ; \,  
\vph_{_{lm}} \to \frac{\vph_{_{lm}}}{r} 
\eeq
one obtains \cite{melnikov}
\br
&&H = \sum_{lm} \fr 1 2 \int_0^\infty dr 
\le\{\p_{lm}^2(r) + r^2 \le[\fr{\pa}{\pa r} \le(\fr{\varphi_{lm} 
(r)}{r}\ri)\ri]^2 \ri. \nn\\
&& \hskip .5in + \le. \fr{l(l+1)}{r^2}~\varphi_{lm}^2(r)\ri\}. 
\la{ham2}
\er
This is nothing but the Hamiltonian of a free scalar field propagating
in flat space-time. This is true for {\it any} fixed value of
$\t$, provided the scalar field is traced over either the region $r
\in (0, r_h]$ or the region $r \in [r_h, \infty)$. 
Note that the black-hole singularity can be entirely avoided for the latter choice.  
Now for evaluating time-independent quantities such as entropy, it
suffices to use the above Hamiltonian (the same cannot be said for
time-dependent quantities).

The approach here differs from that of Ref. \cite{mukoh} where the
authors divide the exterior region $r \geq r_s$ into two by
introducing an hypothetical spherical surface and obtain the
entanglement entropy of that surface. In contrast, we consider the
complete $r \geq r_s$ region and obtain the entropy for the BH horizon.
We discuss the possible extensions in Sec. (\ref{conclu}).

\section{Entanglement entropy for GS and ES}
\label{app:GSES-ent}

For the sake of completeness, we outline the essential steps in the
computation of entanglement entropy for ground state and first excited
state. In the following, we denote all the quantities, viz., wave
function, density matrix etc. by the symbol/suffix $0$ for GS and by
$1$ for the first ES.

\subsection{Ground state}              
\la{ent-gs}

In this case the wave function (\ref{excwavefn1}) reduces to (on setting
$\n_i = 0,~$ for all $i$): 
\bea \la{gs-wavefn}
\psi_0 (x_1, \dots, x_N) &=& \prod_{i=1}^N \cN_i^{(0)} 
\exp \le(- \fr 1 2 k_{Di}^{1/2} {\underbar x}_i^2\ri) \nn\\
&=& \le(\fr{|\O|}{\p^N}\ri)^{1/4} \exp \le[-~ \fr{x^T \O x} 2\ri] \, .
\eea
Let us decompose
\bea \la{Omega}
\O  ~=~ \le( \begin{array}{ll} 
{A} & {B} \\
{B^T} & {C} 
\end{array} \ri) 
\eea
and define
\be
\b ~=~ \frac{B^T A^{-1} B} 2 ~;~~~ \c ~=~ C - \b  \, ,
\ee
where $A$ is an $n \times n$ symmetric matrix, $B$ is an $n \times
(N-n)$ matrix, and $C, \b, \c$ are all $(N-n) \times (N-n)$ symmetric
matrices. The density matrix (\ref{denmatgen1}) reduces to
\cite{sred}:
\be \la{gs-den}
\r_0 (t; t') = \sq{\fr{|\O|}{\pi^{N-n} |A|}} ~\exp \le[- \fr{t^T \c t 
+ t'^T \c t'} 2 ~+~ t^T \b t'\ri].
\ee
In the above, the matrices $B$ and $\b$ are non-zero if and only if
the HOs are interacting. Due to the Gaussian nature of $\r_0 (t; t')$
(\ref{gs-den}), one can make a series of unitary transformations:
\bea \la{diag}
&& V \c V^T = \c_D = \mbox{diag}~,~ {\bar\b} \equiv \c_D^{- 1/2} 
V \b V^T \c_D^{- 1/2}~,\nn\\ 
&& W {\bar\b} W^T = {\bar\b}_D = \mbox{diag}~,~  v \equiv 
W^T \c_D^{1/2} V ~, 
\eea
such that it reduces to a product of $(N-n)$, two HO $(N = 2)$ density 
matrices, in each of which one oscillator $(n = 1)$ is traced over 
\cite{sred}: 

\be \la{gs-den2}
\r_0 (t; t') = \sq{\fr{|\O|}{\pi^{N-n} |A|}} ~\exp \le[- \fr{t^T \c t
+ t'^T \c t'} 2 ~+~ t^T \b t'\ri].
\ee
In the above, the matrices $B$ and $\b$ are non-zero if and only if
the HOs are interacting. Due to the Gaussian nature of $\r_0 (t; t')$
(\ref{gs-den}), one can make a series of unitary transformations:
\bea \la{diag2}
V \c V^T = \c_D = \mbox{diag}~,~~~~ {\bar\b} \equiv \c_D^{- 1/2}
V \b V^T \c_D^{- 1/2}~, \nn \\
W {\bar\b} W^T = {\bar\b}_D = \mbox{diag}~,~~~~  v \equiv
W^T \c_D^{1/2} V ~,
\eea
such that it reduces to a product of $(N-n)$, two HO $(N = 2)$ density
matrices, in each of which one oscillator $(n = 1)$ is traced over
\cite{sred}:
\be \la{gs-den1}
\r_0(t; t') = \sq{\fr{|\O|}{\pi^{N-n} |A|}} ~ \prod_{i=1}^{N-n}
\exp \le[- \fr{v_i^2+v_i'^2}{2} + {\bar\b}_i v_i v_i' \ri],
\ee
where $v_i \in v$ and ${\bar\b}_i \in \bar\b$.
The corresponding entropy is a sum of two HO entropies \cite{sred}:
{\small
\be \la{gs-ent}
S = - \sum_{i=1}^{N-n} \ln[1-\xi_i] + \fr{\xi_i}{1 - \xi_i} \ln\xi_i 
\quad  
\xi_i = \fr{{\bar \b}_i}{1+ \sqrt{1 - {\bar \b}_i^2}} \, .
\ee
}
For the total Hamiltonian $H$, the entanglement entropy is
\be \la{sredresult}
S = \sum_{l=0}^{l_{max}} (2 l + 1) S_l = 
0.3 \le(\fr{R}{a}\ri)^2 \, ,
\ee
where $(2 l + 1)$ is the degeneracy factor that follows from the 
spherical symmetry of the Hamiltonian, $R = a (n + 0.5)$ is the 
radius of the hypothetical spherical surface -- the horizon -- the
DoF inside of which are traced over, and $S_l$ is the entropy for 
a given $l$. Although ideally the upper limit $l_{max}$ should be
infinity, for numerical estimation of the entropy (for a certain 
precision) a very large value of $l_{max}$ is assigned in practice. 
The precision goal $Pr$ is set by demanding that the maximum value 
of $l \equiv l_{max}$ should be such that the percentage change in 
entropy
\be \la{pr-goal}
\vline\fr{S(l_{max}) - S(l_{max} - 5)} {S(l_{max} - 5)}\vline 
\times 100 ~<~ Pr \, .
\ee
For instance, if $Pr = 0.01$, the numerical error in the total entropy 
estimation is less than $0.01\%$.  

In the cases of GCS and SS, it has been shown in \cite{sdshankiES}
that the expression for the total entropy is the same (up to
irrelevant multiplicative factors) as that for GS. 

\subsection{First excited state}         
\la{ent-es}

In this case, we consider one HO is in the excited state while the
rest $N -1$ are in their GS \cite{sdshanki,sdshankiES}. From
Eq.~(\ref{excwavefn1}), we have
\bea \la{es-wavefn}
\psi_1 (x_1 \dots x_N) &=& \sum_{i=1}^N \le(\fr{k_{Di}}{4 \p}\ri)^{1/4} 
\a_i \cH_1 \le(k_{Di}^{1/4} {\underbar x}_i\ri) \nn\\
&& \times~ \exp \le(-\fr 1 2 \sum_{j} k_{D j}^{1/2}~{\underbar x}_j^2\ri).
\eea 
In terms of the pure GS wave function (\ref{gs-wavefn}) this can be
written as
\be \la{es-wavefn1}
\psi_1 (x_1 \dots x_N) = \sqrt{2} \le(\a^T K_D^{1/2} {\underbar x}\ri)
\psi_0 \le(x_1, \dots, x_N\ri),
\ee
where 
\be \la{expcoeff}
\a^T = \le(\a_1, \dots, \a_N\ri)
\ee
are the expansion coefficients and the normalization of $\psi_1$ requires 
$\a^T \a = 1$.

Using Eq.(\ref{denmatgen1}) density matrix can be evaluated and is
given by:
\bea \label{es-den}
\r_1 (t; t') &=&  2 \int \prod_{i=1}^n dx_i 
\le[x'^T \L x\ri] \psi_0 \le(x_i; t\ri) ~\psi_0^\star 
\le(x_i; t'\ri) \nn\\
&=& \k \le[1 - \fr {t^T \L_\c t + t'^T \L_\c t'} 2 
+ t^T \L_\b t'\ri] \r_0 (t; t') \nn\\
\eea
where $\L$ is a $N \times N$ matrix defined by
\be \la{lambda}
\L =  U^T~K_D^{1/4}~\a~\a^T~K_D^{1/4}~U \equiv
\le( 
\begin{array}{cc} 
\L_A & \L_B \\
\L_B^T & \L_C 
\end{array}
\ri) \, ,
\ee
$\L_A$ is an $n \times n$ symmetric matrix; $\L_B$ is an $n \times
(N-n)$ matrix; $\L_C$ is an $(N-n) \times (N-n)$ symmetric matrix,
and $\k = \mbox{Tr} (\L_A A^{-1})$. The $(N-n) \times (N-n)$ matrices
$\L_\b$ and $\L_\c$ are given by
\bea \la{lambdabetgam}
\L_\b &=& \fr 1 {\k} \le(2 \L_C ~-~ \L_B^T A^{-1} B \ri. \nn\\
&& \hskip .2in \le. -~ B^T A^{-1} \L_B ~+~ B^T A^{-1} \L_A A^{-1} 
B\ri) \nn\\
\L_\c &=& \fr 1 {\k} \le(2 \L_B^T A^{-1} B ~-~ B^T A^{-1} \L_A 
A^{-1} B\ri).
\eea 
$\L_\b$ is symmetric, while $\L_\c$ is not necessarily symmetric due
to the presence of the first term on the right hand side.

In general, unlike the GS density matrix $\r_0$ (\ref{gs-den}), the ES
density matrix $\r_1$ (\ref{es-den}) cannot be factorized into $(N-n)$
two HO density matrices. However, $\r_0$ is a Gaussian that attenuates
virtually to zero beyond its few sigma limits. Therefore, if
\bea \la{smallness}
\e_1 \equiv t_{max}^T  \L_\b  t_{max}  \ll 1~,~ 
\e_2 \equiv t_{max}^T  \L_\c  t_{max}~ \ll 1 
\eea
where 
\be \la{tmax}
t_{max}^T = \le(\fr{3 (N-n)}{\sq{2 \mbox{Tr}(\c - \b)}} \ri)
\le(1,1,\dots \ri)~
\ee
corresponding to $3\sigma$ limits of the Gaussian inside $\r_0$, then
one may approximate  
\bea
1 &-& \fr {t^T \L_\c t + t'^T \L_\c t'} 2 + t^T \L_\b t' \nn \\
&\approx& \exp \le[- \fr {t^T \L_\c t + t'^T \L_\c t'} 2
+ t^T \L_\b t' \ri].
\eea
As such, with the following shift of parameters
\bea \la{es-shift}
\b' \equiv \b + \L_\b ~,~~ \c' \equiv \c + \L_\c
\eea
the approximated ES density matrix is also a Gaussian
\bea \la{es-den1}
\rho_1 (t; t') \approx \k \exp \le[-\fr {t^T \c' t 
+ t'^T \c' t'} 2 + t^T \b' t'\ri].
\eea
This can be factorized once again into two HO density matrices, and the 
associated entanglement entropy can be computed. For the following set 
of values: $N = 300, ~n = 100 - 200, ~o = 10 - 50$, $o$ being the number 
of last non-vanishing entries in the vector $\a^T$, i.e., $\a^T = 
\le(1/\sq{o}\ri) (0, \cdots, 0, {1, \cdots, 1})$, the entropy computation
is done numerically (using MATLAB) in \cite{sdshankiES}. The precision 
setting in the computation had been $0.1 \%$ and the criteria (\ref{smallness})
is satisfied for the above choice of the parameters $N, n$ and $o$. The 
results show that the ES entropy scales as a power of the area. The power, 
however, is always less than unity (for any $o > 0$) and is lesser and lesser, 
the higher the value of $o$. The AL is thus always violated for the chosen
range of values of $n (= 100 - 200)$. The other 
interesting observation made in \cite{sdshankiES} is the shifting of 
peaks in the variation of the partial waves $(2l+1)S_l$ with $l$ for ES 
as compared to the case for GS.


\end{document}